\documentclass[1p]{elsarticle}

\usepackage{lineno,hyperref}
\modulolinenumbers[5]

\makeatletter
\def\ps@pprintTitle{%
 \let\@oddhead\@empty
 \let\@evenhead\@empty
 \def\@oddfoot{\centerline{\thepage}}%
 \let\@evenfoot\@oddfoot}
\makeatother

\journal{}

\usepackage{hhline}
\usepackage{multirow}
\usepackage{multicol}
\usepackage{tabularx}
\usepackage{amsmath, amssymb}  
\usepackage{bm}  
\usepackage{float} 
\usepackage{xcolor} 
\usepackage{caption}
\usepackage{subcaption}
\usepackage{relsize}
\usepackage{xspace}
\usepackage{svg}

\usepackage{textcomp}  

\newcommand{\Rplus}{\protect\hspace{-.1em}\protect\raisebox{.35ex}{\smaller{\smaller\textbf{+}}}}
\newcommand{\Cpp}{\mbox{C\Rplus\Rplus}\xspace}

\begin{document}

\begin{frontmatter}

\title{\textit{giotto-ph}: A Python Library for High-Performance Computation of Persistent Homology of Vietoris--Rips Filtrations}

\author[reds]{Julián Burella Pérez}
\ead{julian.burellaperez@heig-vd.ch}
\fntext[myfootnote]{The first two authors contributed equally to this work.}
\author[reds]{Sydney Hauke}
\ead{sydney.hauke@heig-vd.ch}
\author[epfl]{Umberto Lupo\corref{mycorrespondingauthor}}
\cortext[mycorrespondingauthor]{Corresponding author}
\ead{umberto.lupo@epfl.ch}
\author[l2f]{Matteo Caorsi}
\ead{m.caorsi@l2f.ch}
\author[reds]{Alberto Dassatti}
\ead{alberto.dassatti@heig-vd.ch}
\address[reds]{HEIG-VD, HES-SO, Route de Cheseaux 1, Yverdon-les-Bains, Switzerland}
\address[epfl]{EPFL, Route Cantonale, Lausanne, Switzerland}
\address[l2f]{L2F SA, Rue du centre 9, Saint-Sulpice, Switzerland}

\begin{abstract}
We introduce \textit{giotto-ph}, a high-performance, open-source software package for the computation of Vietoris--Rips barcodes.  \textit{giotto-ph} is based on Morozov and Nigmetov's lockfree (multicore) implementation of Ulrich Bauer's \textit{Ripser} package.  It also contains a re-working of the \textit{GUDHI} library's implementation of Boissonnat and Pritam's \textit{Edge Collapser}, which can be used as a pre-processing step to dramatically reduce overall run-times in certain scenarios.  Our contribution is twofold: on the one hand, we integrate existing state-of-the-art ideas coherently in a single library and provide Python bindings to the \Cpp code.  On the other hand, we increase parallelization opportunities and improve overall performance by adopting more efficient data structures.  Our persistent homology backend establishes a new state of the art, surpassing even GPU-accelerated implementations such as \textit{Ripser++} when using as few as 5--10 CPU cores.  Furthermore, our implementation of \textit{Edge Collapser} has fewer software dependencies and improved run-times relative to \textit{GUDHI}'s original implementation.

\end{abstract}


\end{frontmatter}


\section{Introduction}
\label{sec:intro}
In recent years, \emph{persistent homology} (PH) (see e.g.\ \cite{ghrist2007barcodes, edelsbrunner2008persistent, edelsbrunner2014persistent, oudot2015persistence, chazal2016structure, perea2018brief, carlsson2019persistent, nanda2021computational} for surveys) has been a key driving force behind the ever-increasing adoption of topological approaches in a wide variety of computational contexts, such as geometric inference \cite{edelsbrunner2014short, boissonnat2018geometric}, signal processing \cite{robinson2014topological, perea2015sliding}, data visualization \cite{tierny2018topological}, and, more generally, data analysis \cite{carlsson2009topology, chazal2021introduction} and machine learning \cite{hensel2021survey}.\footnote{As a matter of fact, insights from the theory of persistent homology and \emph{persistence modules} have also proved helpful in pure mathematics, e.g.,\ in symplectic topology/geometry \cite{polterovich2020topological}.}  Among the main invariants described by this theory, the (\emph{persistence}) \emph{barcode} \cite{frosini1990shapes, frosini1992measuring, barannikov1994morse, robins1999approximations, edelsbrunner2000simplification, zomorodian2005computing} has attracted the most attention due to a) its ability to track the appearance and disappearance of topological features in data throughout entire ranges of parameters, b) its succinct nature and ease of representation, as it simply consists of a (typically small) collection of intervals of the real line, c) its provable robustness under perturbations of the input data \cite{damico2003optimal, cohen-steiner2007stability}, and d) its amenability to computation and algorithmic optimization, as demonstrated by the large number of existing implementations -- see Sect.\ 1 in \cite{bauer2021ripser} for a review, and \cite{aggarwal2021dory, vonbromssen2021computing} for recent entries not mentioned there.

Despite these successes, the computation of barcodes remains a challenge when dealing with large datasets and/or with high-dimensional\footnote{In this paper, we use the word ``dimension'' as a synonym for ``degree'' when describing homology classes and therefore bars in a persistence barcode.} topological features.  We now explain why this is the case.  The input to any barcode computation is a growing, one-parameter family of combinatorial objects, called a \emph{filtration} or a \emph{filtered complex}.  Filtrations consist of \emph{cells} with assigned integer \emph{dimensions} and values of the filtration-defining parameter, as well as \emph{boundary} (resp.\ \emph{co-boundary}) relations (mappings) between $k$-dimensional cells and ($k-1$)-dimensional [resp.\ ($k+1$)-dimensional] ones.\footnote{We refer the reader to any of the aforementioned surveys of PH for rigorous definitions.}  Arguably, the most common examples of filtrations in applications concern \emph{simplicial} complexes, in which case the cells are referred to as \emph{simplices}, and $k$-dimensional simplices consist of sets of $k + 1$ points from a common vertex set $V$ -- for instance, a $0$-simplex $\{v\}$ is one of the vertices $v \in V$, while a $1$-simplex $\{v, w\} \subseteq V$ can be thought of as an edge connecting vertices $v$ and $w$.  These are the filtrations of interest in the present paper.  In particular, we focus on the \emph{Vietoris--Rips} (VR) (resp.\ \emph{flag}) filtration of a finite metric space (resp.\ undirected graph with vertex and edge weights), in which simplices are arbitrary subsets of the available points (resp.\ vertices) and their filtration values are set to be their diameters (resp.\ the maximum weights of all vertices and edges they contain, with absent edges being given infinite filtration value).

Several simplicial filtrations of interest in applications, and the VR filtration chiefly among these, quickly become very large as their defining parameter increases (and hence more and more simplices are included in the growing complex).  At the heart of all algorithms for computing PH barcodes lies the reduction of \emph{boundary} or \emph{co-boundary matrices} indexed by the full set of simplices in the filtration; the available reduction algorithms have asymptotic space and time complexities which are polynomial in the total number $N$ of simplices.  In the case of the VR filtration of an input metric space $\mathcal{M}$, if one is interested in computing the barcode up to and including homology dimension $D$, then $N = \sum_{k=0}^{D + 1} \binom{|M|}{k}$.  For sizeable datasets, this combinatorial explosion leads to a staggering number of elementary row or column operations (as well as memory) required to distil the desired barcode.  PH computation for many other simplicial filtrations constructed from point clouds, finite metric spaces, or graphs are also ultimately limited by similar considerations.

\subsection{Related work}
\label{sec:sota}

To the best of our knowledge, at the time of writing \textit{Ripser}\footnote{\url{https://github.com/Ripser/ripser}, retrieved 1 August 2021.} \cite{bauer2021ripser} is the \emph{de facto} state of the art and reference for computing VR persistence barcodes on CPUs.  \textit{Ripser} uses multiple known optimizations like \emph{clearing} \cite{chen2011persistent} and \emph{cohomology} \cite{desilva2011dualities}.  Furthermore, it makes use of other performance-oriented ideas, such as the implicit representation of the (co)boundary and reduced (co)boundary matrices, and the \emph{emergent/apparent pairs} optimizations (we refer to \cite{bauer2021ripser} for definitions and details).  At the time of writing, the latest version of \textit{Ripser} is \textit{v1.2}.\footnote{Release date: 25 February 2021.}  In that version, to the emergent pairs optimization in use until that point was added an optimization based on apparent pairs.

Although \textit{Ripser v1.2} is arguably the fastest existing code for computing VR barcodes in a sequential (i.e.,\ single CPU core) setting, it has no parallel capabilities.  Overcoming this limitation is possible as two recent lines of work \cite{morozov2020towards, zhang2020gpuaccelerated} demonstrate.  Based on a ``pairing uniqueness lemma'' proved by Cohen-Steiner \textit{et al.}\ \cite{cohen-steiner2006vines}, Morozov and Nigmetov \cite{morozov2020towards} observe that the reduction of the (co)boundary matrix can, in fact, be performed out of order as long as column additions are always performed left to right.\footnote{``Left to right'' here is meant relative to the filtration order.}  Therefore -- these authors suggest -- any column reduction can be efficiently performed in parallel provided adequate synchronisation is used.  A functional proof of concept of this idea as applied to a now superseded\footnote{In particular, based on \textit{Ripser v1.1} and hence not using apparent pairs, cf.\ Figure \ref{fig:lib}.} version of \textit{Ripser} was put in the public domain in June 2020 \cite{morozov2020lock}, but has not led to a distributable software package.  A generic implementation of the ideas in \cite{morozov2020towards}, not tied to Vietoris--Rips filtrations and instead designed to make lock-free reduction possible on any (co)boundary matrix, has recently been published as the \textit{Oineus} library \cite{nigmetov2020oineus}.  Although optimizations such as clearing and implicit matrix reduction appear to have been implemented there, the code and performance are not optimized for Vietoris--Rips filtrations, and in particular no implementation of the emergent \emph{or} apparent pair optimization is present there at this time.\footnote{At the time of writing, this library is in version 1.0.  We were not aware of its existence during the development of our code.}

Zhang \emph{et al.}'s \textit{Ripser++} \cite{zhang2020gpuaccelerated} implements the idea of finding apparent pairs in parallel on a GPU to accelerate the computation of VR barcodes.  Despite this, \textit{Ripser++} is not fully parallel.  For each dimension to process, it divides the computation into three sub-tasks: ``\textit{filtration construction and clearing}'', ``\textit{finding apparent pairs}'' and ``\textit{sub-matrix reduction}''.  The last of these steps is not parallel and it is executed on the CPU.  Accordingly to Amdahl’s law,\footnote{\url{https://en.wikipedia.org/wiki/Amdahl's\_law}} this processing sequence is expected to yield only diminishing returns when augmenting the number of parallel resources.  Although performance gains have been demonstrated in \cite{zhang2020gpuaccelerated} -- particularly when using high-end GPUs -- there is room for extending parallelism to the third sub-task above, which we try to harvest in this work by integrating the aforementioned ideas from \cite{morozov2020towards}.

All implementations presented in this subsection (barring \cite{nigmetov2020oineus}, which also provides some Python bindings) are written in low-level languages (\Cpp, CUDA).  However, researchers today make wide use of higher-level languages -- Python, for instance, is the dominant one in several fields.  It is therefore natural that libraries have been developed to couple Python's ease of use with the high performance provided by these pieces of code.  \textit{Ripser.py} \cite{ctralie2018ripser} is probably the most notable of these implementations, providing an intuitive interface for VR filtrations wrapping \textit{Ripser} at its core.  The authors of the library forked the original \textit{Ripser} implementation and added support for non-zero birth times, as well as the possibility to compute and retrieve cocycles.

Meanwhile, in 2020, Boissonnat and Pritam presented a new algorithm they called \textit{Edge Collapser} (EC) \cite{boissonnat2020edge}.  Independently of the code used to compute barcodes, EC can be used as a pre-processing step on any flag filtration to remove ``redundant'' edges -- and modify the filtration values of some others -- while ensuring that the flag filtration obtained from the thus ``sparsified'' weighted graph has the same barcode as the original filtration.  Although it introduces an initial overhead, pre-processing by EC can dramatically improve the end-to-end run-time for barcode computation by greatly reducing the complexity of the downstream reduction steps.  As reported by those authors, this is especially true when one wishes to compute barcodes in high homology dimensions, and/or when one is dealing with large datasets.  An implementation of EC has already been integrated into the \textit{GUDHI} library \cite{gudhi:urm, gudhi:Collapse}.

\subsection{Our contribution}

In this context, we present \textit{giotto-ph},\footnote{\textit{giotto-ph} is available at \url{https://github.com/giotto-ai/giotto-ph}.} a Python package built on top of a \Cpp backend that computes PH barcodes for VR filtrations on the CPU.  To the best of our knowledge, this is the first package fully integrating the three ideas described in Section \ref{sec:sota} (lock-free reduction, parallelized search for apparent pairs, edge collapses) in a single portable, easy-to-use library.  We remark that, after the release of our code and of the first version of this paper, we learned about a very recent thesis \cite{tulchinskii2021fast} and associated publicly available code,\footnote{\url{https://github.com/ArGintum/PersistenceHomology}, retrieved 1 August 2021.} in which a very similar program to ours has been carried out, though the apparent pairs optimization and support for coefficients in finite prime fields other than $\mathbb{F}_2$ are among the features still missing from that package.

When developing \textit{giotto-ph} we focused on increasing execution speed throughout.  In particular:

\begin{enumerate}

  \item We built on the ideas for parallel reduction presented in \cite{morozov2020towards} and on the prototype implementation described in \cite{morozov2020lock}, and improved execution speed and resource usage by implementing custom lock-free hash tables and a thread pool.
  
  \item Similarly to \textit{Ripser++}, we implemented a parallel version of the apparent pairs optimization, thus far only present in serial form in \textit{Ripser v1.2}.
  
  \item We re-implemented the EC algorithm to increase its execution speed compared to \cite{gudhi:Collapse}.  The simple observation that the well-known \emph{enclosing radius} optimization is applicable to EC is shown here to lead to even larger improvements.
  
\end{enumerate}

Our results show that our code is often $1.5$ to $2$ times (and, in one example, almost $8$ times) faster than \cite{morozov2020towards} and able to beat \textit{Ripser++} \cite{zhang2020gpuaccelerated}, the current state-of-the-art GPU implementation, while running only on CPU and with as few as 5--10 cores.

Finally, \textit{giotto-ph} owes some architectural decisions to \textit{Ripser.py} \cite{ctralie2018ripser} -- in particular, the support for node weights.  At the level of the Python interface, our main contribution is supporting \emph{weighted Rips filtrations} -- in particular, the \emph{distance-to-measure}--based filtrations described in \cite{anai2020dtmbased}.

Thanks to its reduced memory usage and shorter run-times, we hope that \textit{giotto-ph} will enable researchers to explore larger datasets, and in higher homology dimensions, than  ever before.

\subsection{Structure of the paper}

In Section \ref{sec:implementation} we describe the package implementation, detailing optimizations adopted to increase speed and portability. In Section \ref{sec:experiments}, we show how our library compares to other implementations. We also report performance when increasing the maximum homology dimension computed, observing that in that regime we are shifting the limiting factor from memory towards the enumeration of simplices, currently limited to $2^{64}$. Finally, in Section \ref{sec:conclusion} we draw conclusions and sketch future research directions.

\section{Implementation}
\label{sec:implementation}

\begin{figure}[H]
    \begin{center}
        \includegraphics[width=\linewidth]{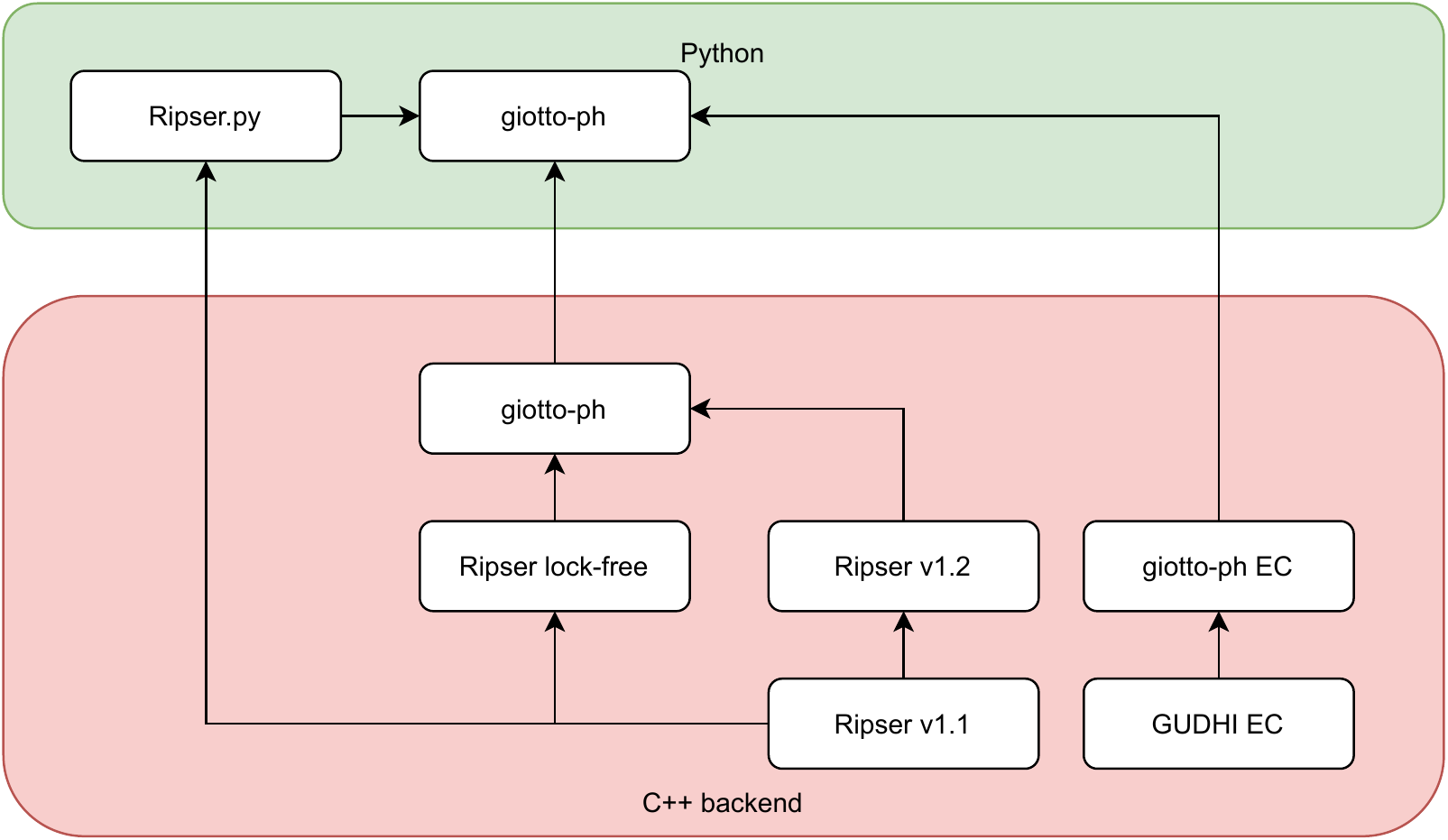}
    \end{center}
    \caption{\textit{giotto-ph} consists of a \Cpp backend and a Python frontend. The Python interface is based on \textit{Ripser.py}~\cite{ctralie2018ripser} (see section \ref{sec:python} for details). The figure also shows the inheritance of \textit{giotto-ph}'s \Cpp backend from pre-dating implementations.}
    \label{fig:lib}
\end{figure}


\textit{giotto-ph} is a library dedicated to the efficient computation of PH of VR filtrations (see Section \ref{sec:intro}). It inherits and extends ideas and code from many sources; Figure \ref{fig:lib} gives a visual representation of the most important ones among them. Our aim with \textit{giotto-ph} is to provide an alternative to the excellent \textit{Ripser.py} library, retaining several of the latter's advantages, namely portability and ease of use,\footnote{Currently, \textit{giotto-ph} does not support retrieving cocycles, but there are no substantial challenges, in principle, to the addition of this feature in the near future.} while replacing the \Cpp backend with a new parallel and higher-performance version.  

\subsection{\Cpp backend}
\label{subsec:Cpp_backend}



The implementation of \textit{giotto-ph}'s backend parallel algorithm is heavily inspired by \cite{morozov2020lock}, a functional proof of concept of \cite{morozov2020towards}.  Starting from \cite{morozov2020towards}, we replaced the main data structure and the threading strategy to minimize the overhead introduced by adding parallelism. Furthermore, we introduced the apparent pairs approach, in its parallel form, to harvest its benefits in shortening run-times: a decreased number of columns to reduce \cite{bauer2021ripser} and an additional early stop condition when enumerating cofacets.


The main data structure of the algorithm described in \cite{morozov2020towards} is a lock-free hash table. A lock-free hash table is a concurrent hash map where concurrent operations do not make use of synchronization mechanisms such as mutexes. Instead, a lock-free hash table relies on atomic operations for manipulating its content; in particular, insertion is carried out by a mechanism called compare-and-swap (CAS).  After benchmarking a few performance-oriented alternatives offering portability for most available compilers for Linux, Mac OS X, and Windows systems, we created a custom hash map adapted to the needs of the core matrix reduction algorithm, using the ``Leapfrog'' implementation in the open-source \textit{Junction} library.\footnote{\url{https://github.com/preshing/junction}}

As previously mentioned, we also adopted a different threading strategy: a \emph{thread pool}.\footnote{With optional CPU pinning option.}  A thread pool is a design pattern in which a ``pool" of threads is created up front when the program starts, and the same threads are reused for different computations during the program's life span.  This approach enables better amortization of the cost of the short-lived threads used in \cite{morozov2020lock}, where one thread is created whenever needed and destroyed at the end of its computation task.  Table \ref{tab:pool} compares the running time of a solution based on our thread pool with the former approach. The run-time improvements are highly dataset dependent, but always measurable in the considered scenarios.

The final component in our \Cpp backend is a rewriting of the EC algorithm (see Section \ref{sec:sota}), implemented so far only in the \textit{GUDHI} library \cite{gudhi:Collapse}.  Our implementation focuses on performance and removes the dependencies on the \textit{Boost} \cite{BoostLibrary} and \textit{Eigen} \cite{eigenweb} libraries.  \textit{giotto-ph}'s EC is more than 1.5 times faster than the original version as reported in Table \ref{tab:collapser}.  It also supports weighted graphs with arbitrary (possibly non-positive) edge weights as well as arbitrary node weights.  Improvements were achieved mainly by reworking data structures, making the implementation more cache-friendly, and directly iterating over data without any transformation, hence reducing the pressure on the memory sub-system.


\begin{table}[h!]
\centering
\begin{tabular}{l|r|r|r|r|}
\cline{2-5} & \multicolumn{4}{c|}{\textbf{giotto-ph backend}} \\ \cline{2-5} 
            & \multicolumn{2}{c|}{\textbf{no thread pool}} & \multicolumn{2}{c|}{\textbf{thread pool}} \\ \hline
\multicolumn{1}{|l|}{{\textbf{dataset}}} & \multicolumn{1}{c|}{$\bm{N = 8}$} & \multicolumn{1}{c|}{$\bm{N = 48}$} & \multicolumn{1}{c|}{$\bm{N = 8}$} & \multicolumn{1}{c|}{$\bm{N = 48}$} \\ \hline
\multicolumn{1}{|l|}{\texttt{sphere3}}    & {0.4} & {0.4}  & {0.4}  & 0.38 \\ \hline
\multicolumn{1}{|l|}{\texttt{dragon}}     & {1.2} & {1.2}  & {1.3}  & 1.3 \\ \hline
\multicolumn{1}{|l|}{\texttt{o3\_1024}}   & {0.4} & {0.18} & {0.4}  & 0.17 \\ \hline
\multicolumn{1}{|l|}{\texttt{random16}}   & {0.9} & {0.4}  & {0.9}  & 0.24 \\ \hline
\multicolumn{1}{|l|}{\texttt{fractal}}    & {0.9} & {0.35} & {0.9}  & 0.34 \\ \hline
\multicolumn{1}{|l|}{\texttt{o3\_4096}}   & {6.9} & {2.7}  & {6.9}  & 2.6 \\ \hline
\multicolumn{1}{|l|}{\texttt{torus4}}     & {19}  & {14.7} & {19.1} & 14.3 \\ \hline
\end{tabular}
\caption{Running times, expressed in seconds, with and without the thread pool.  $N$ denotes the number of threads used.  All information regarding the datasets presented here are described in section \ref{sec:experiments} and summarized in Table \ref{tab:datasets}.}
\label{tab:pool}
\end{table}


\subsection{Python Interface}
\label{sec:python}
Our Python interface is based on \textit{Ripser.py} \cite{ctralie2018ripser}.  While it lacks some of \textit{Ripser.py}'s features, such as the support for ``greedy permutations'' and for retrieving cocycles, it introduces the following notable improvements:

\begin{description}

\item[Support for Edge Collapser] 
 EC is disabled by default because it is expected and empirically confirmed that, unless the data is large and/or the maximum homology dimension to compute is high, the initial run-time overhead due to EC is often not compensated for by the resulting speed-up in the downstream reduction steps (see end of Section \ref{sec:sota}).  However, users can easily enable it by means of the \texttt{collapse\_edges} optional argument.  In Table \ref{tab:higher} we will show the difference in run-times when this option is active.  See also ``Support for enclosing radius'', below.

\item[Support for enclosing radius] The \emph{(minimum) enclosing radius} of a finite metric space is the radius of the smallest enclosing ball of that space.  Its computation, starting from a distance matrix, is trivial to implement and takes negligible run-time on modern CPUs.  Above this filtration value, the Vietoris--Rips complex becomes a cone, and hence all homology groups are trivial.  Hence, simplices with higher filtration values than the enclosing radius can be safely omitted from the enumeration and matrix reduction steps, without changing the final barcode.  When the enclosing radius is considerably smaller than the maximum distance in the data, this can lead to dramatic improvements in run-time and memory usage, as observed in \cite{henselmanpetrusek2020matroids}.\footnote{For instance, the barcode computation for the \texttt{random16} dataset (see Table \ref{tab:datasets}) up to dimension $7$ would not be completed after two hours without the enclosing radius optimization; with it, the run-time drops to seconds.  Not all datasets can be expected to witness equally impressive improvements, but the cost of computing the enclosing radius is trivial compared to the computation of PH.}  Unless the user specifies a threshold, both \textit{Eirene} \cite{henselmanghristl6} and \textit{Ripser} make use of the enclosing radius optimization, and the same is true in \textit{giotto-ph}, where the enclosing radius computation is implemented in Python using highly optimized \textit{numpy} functions.  An element of novelty in our interface is that, when both the enclosing radius is computed and EC is enabled, the input distance matrix/weighted graph is thresholded \emph{before} being passed to the EC backend.  As we experimentally find and report in Section \ref{subsec:Collapser}, on several datasets this can lead to substantial run-time improvements for the EC step.


\item[Weighted VR filtrations] While standard stability results for VR barcodes \cite{cohen-steiner2007stability, chazal2009proximity} guarantee robustness to small perturbations in the data, VR barcodes are generally \emph{unstable} with respect to the insertion or deletion of even a single data point.  Thus, even relatively small changes in the local density can greatly affect the resulting barcodes, rendering the vanilla VR persistence pipeline very vulnerable to statistical outliers. Distance-to-measure (DTM) based filtrations \cite{anai2020dtmbased} address this issue by re-weighting vertices and distances according to the local neighbourhood structure. The user can toggle DTM-based reweighting (or more general reweightings) by appropriately setting the optional parameters \texttt{weights} and \texttt{weight\_params}.

\item[\textit{pybind11}\footnotemark \ bindings]\footnotetext{\url{https://github.com/pybind/pybind11}} We added support for and used \textit{pybind11} instead of \textit{Cython}\footnote{\url{https://cython.org/}} for creating Python bindings.  In our experience, it is easier to use without compromising performance. Furthermore, it is already used for the bindings in the \textit{giotto-tda} library \cite{tauzin2021giottotda}, our sibling project.  The presence of Python bindings as well as the portability on different operating systems, namely Linux, Mac OS X, and Windows, have been two of our core objectives to facilitate the adoption of our library.
\end{description}

\section{Experimental results}
\label{sec:experiments}
All experiments presented in this paper were performed on a machine running Linux CentOS 7.9.2009 with kernel 5.4.92, equipped with two Intel\textsuperscript{\textregistered} XEON\textsuperscript{\textregistered} Gold 6248R (24 physical cores each) and a total of 128 GB of RAM. 


We present measures on the datasets of Table \ref{tab:datasets} because they are publicly available, and they are used in publications \cite{Otter_2017, bauer2021ripser} describing established algorithms, making them a representative benchmark set and facilitating comparisons among competing solutions. All datasets are stored as point clouds. When the \texttt{threshold} parameter is empty, the tests report run-times with the enclosing radius option active. The \texttt{dim} parameter corresponds to the maximum dimension for which we compute PH, and the \texttt{coeff} parameter corresponds to the prime field of coefficients (in our tests, this is always $\mathbb{F}_2$).


\begin{table}[H]
            \centering
\begin{tabular}{|l|r|l|r|r|}
\hline
\multicolumn{1}{|c|}{\textbf{dataset}} & \multicolumn{1}{c|}{\textbf{size}} & \multicolumn{1}{c|}{\textbf{threshold}} & \multicolumn{1}{c|}{\textbf{dim}} & \multicolumn{1}{c|}{\textbf{coeff}} \\ \hline
\texttt{sphere3} & 192   &                                         & 2 & 2 \\ \hline
\texttt{dragon} & 2000   &                                         & 1 & 2 \\ \hline
\texttt{o3\_1024} & 1024 & \multicolumn{1}{r|}{1.8}                & 3 & 2 \\ \hline
\texttt{random16} & 50   &                                         & 7 & 2 \\ \hline
\texttt{fractal} & 512   &                                         & 2 & 2 \\ \hline
\texttt{o3\_4096} & 4096 & \multicolumn{1}{r|}{1.4}                & 3 & 2 \\ \hline
\texttt{torus4}  & 50000 & \multicolumn{1}{r|}{0.15}               & 2 & 2 \\ \hline
\end{tabular}
\caption{Datasets used for bechmarking.  ``Size'' means the number of points in the dataset.}
\label{tab:datasets}
 \end{table}


\subsection{Comparison with state-of-the-art algorithms}
In this section we compare our implementation with other implementations of PH for VR filtrations that use an approach similar to Ripser.  We do not directly compare with other existing libraries which adopt different approaches, like \textit{GUDHI} \cite{gudhi:urm} and \textit{Eirene} \cite{henselmanghristl6}, because from \cite{bauer2021ripser} it is evident that \textit{Ripser} is always faster. 

\begin{figure}[H]
    \begin{center}
        \includegraphics[width=0.8\linewidth]{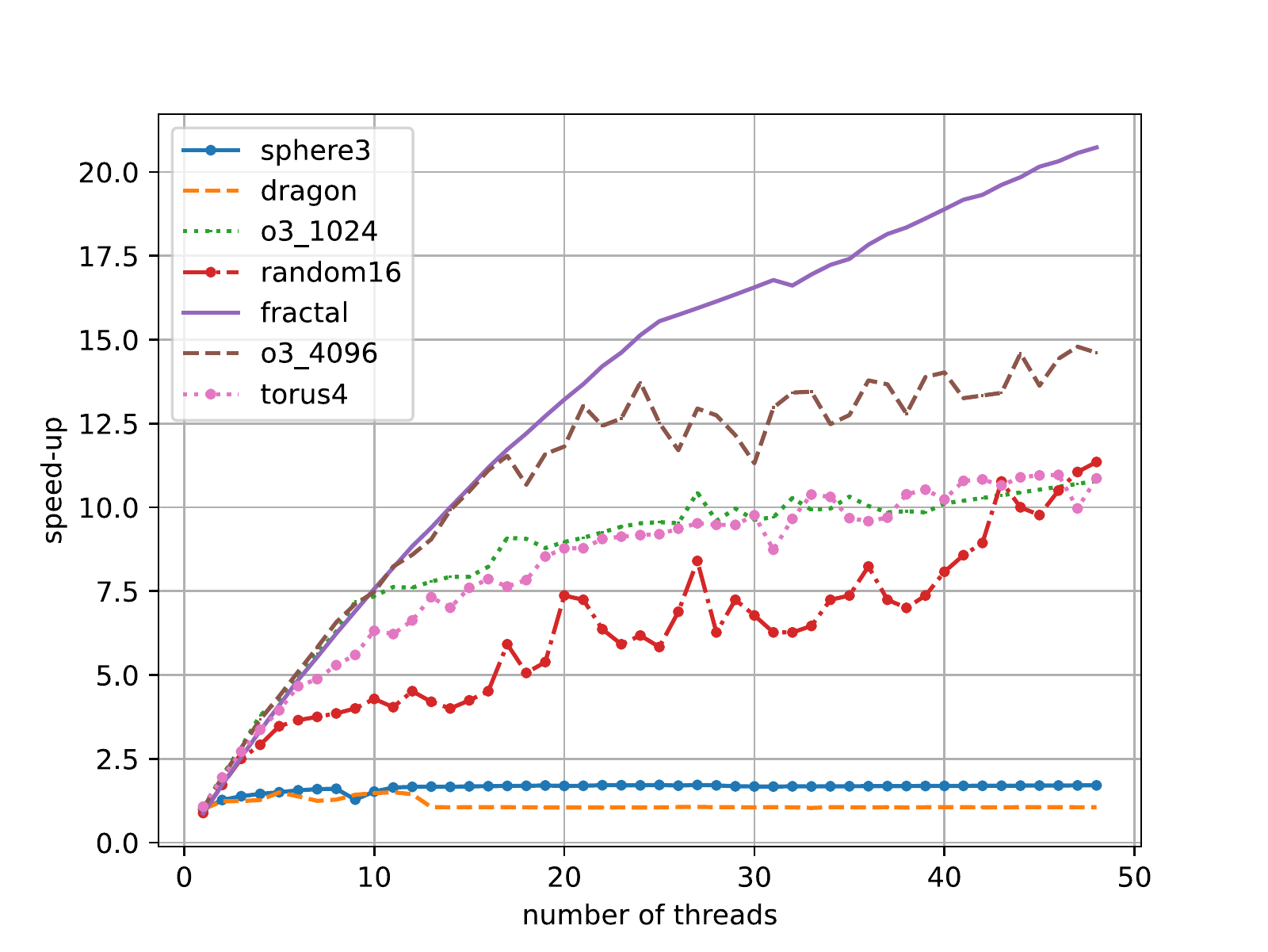}
    \end{center}
    \caption{Speed-up of \textit{giotto-ph} compared to \textit{Ripser v1.2}. \textit{giotto-ph} is always faster than \textit{Ripser v1.2} when using more than one thread.  For the \texttt{dragon} dataset, there is little speed-up in general, and virtually no speed-up ($\sim 1.05$) with 13 threads or above; in that case we compute homology only up to dimension $1$ and the cost of setting up the parallel element of the library is non-zero.}
    \label{fig:1.2}
\end{figure}

Figure \ref{fig:1.2} compares the \textit{giotto-ph} backend and \textit{Ripser v1.2}.  When the computation of the filtration is very fast, due to the reduced number of points or the low dimension of the computation, there is marginal or no benefit in adopting our parallel approach.


\begin{figure}[H]
    \begin{center}
        \includegraphics[width=0.8\linewidth]{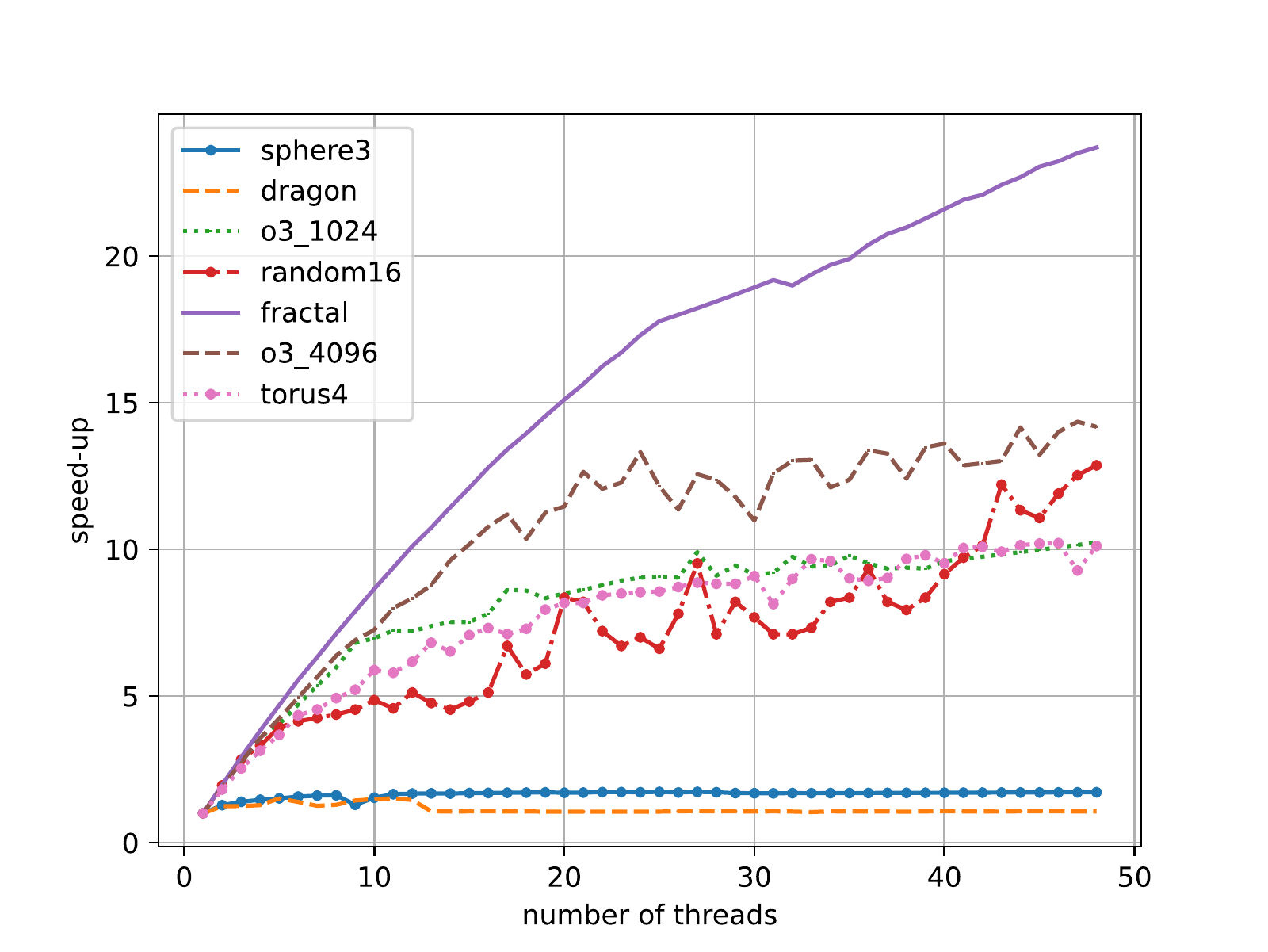}
    \end{center}
    \caption{Scaling of \textit{giotto-ph} when increasing the number of threads. This figure is similar to Figure \ref{fig:1.2} because in a single thread configuration \textit{giotto-ph} performs very similarly to \textit{Ripser v1.2}.}
    \label{fig:scaling}
\end{figure}

Figure \ref{fig:scaling} shows the scaling of \textit{giotto-ph} when increasing the number of worker threads.  Scaling is different for each dataset due to the variable number of apparent and emergent pairs as well as the dimension parameter used.  Observe that the larger the number of points in the dataset, the better the scaling.  Similar effects are visible for datasets in which when the maximal homology dimension to compute is higher.

As seen above for run-times versus \textit{Ripser v1.2}, for scaling too the least favourable results are obtained on datasets such as \texttt{sphere3} and \texttt{dragon}, in which both the size and maximum homology dimension to compute are small.


\begin{figure}[H]
    \begin{center}
        \includegraphics[width=0.8\linewidth]{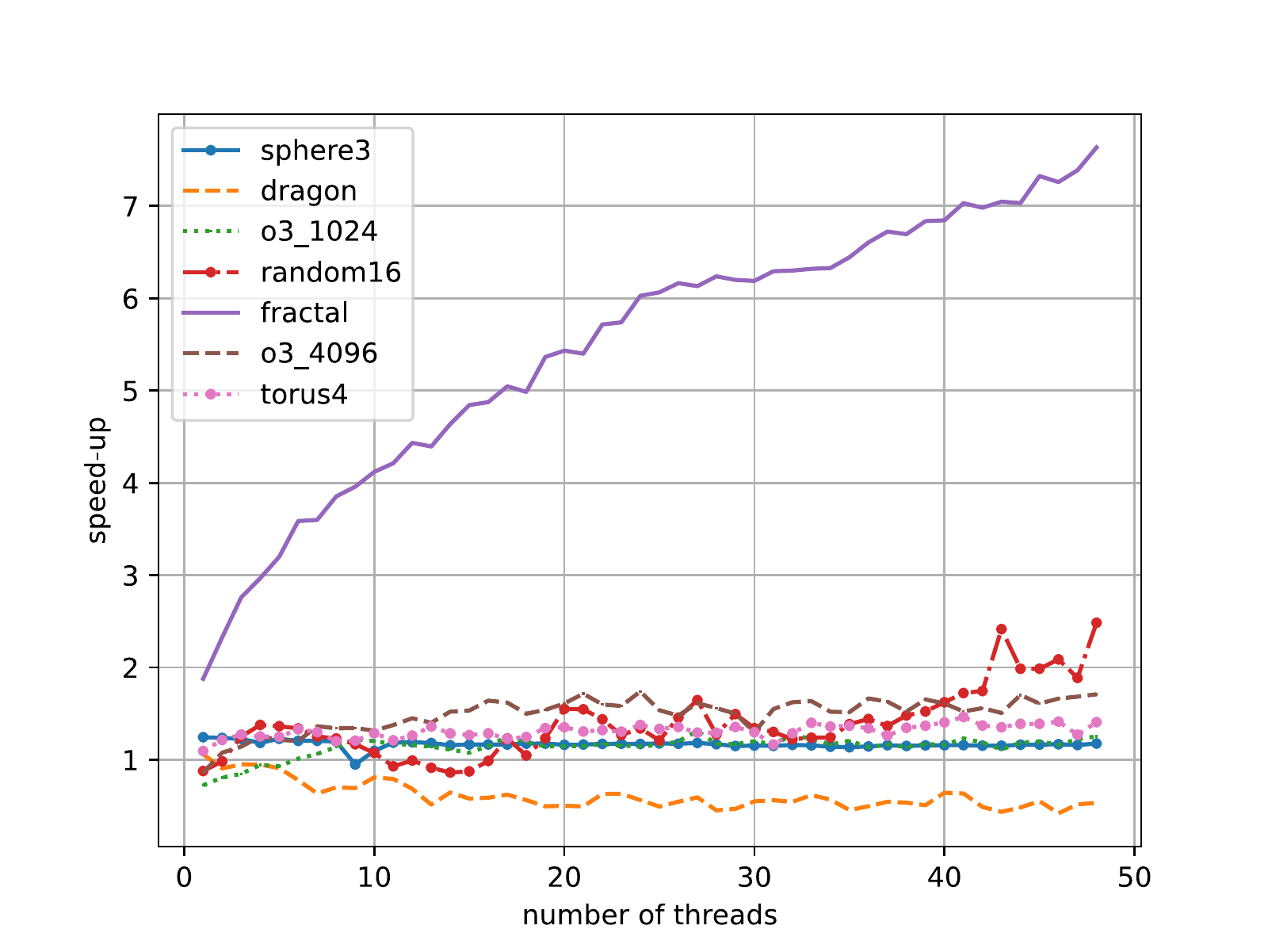}
    \end{center}
    \caption{Speed-up of \textit{giotto-ph} compared to the implementation in \cite{morozov2020lock}.  \textit{giotto-ph} is faster in general, but with the \texttt{fractal} dataset the speed-up is larger -- almost a factor of $8$ when $48$ threads are used. This phenomenon is explained by the large number of apparent pairs in this specific dataset.  On the other hand, performance on \texttt{dragon} is worse as explained in the main body of text.}
    \label{fig:moro}
\end{figure}

According to our measurements reported in Figure \ref{fig:moro}, our implementation outperforms Morozov and Nigmetov's proof-of-concept implementation \cite{morozov2020lock} in most cases, and most noticeably when the number of parallel resources increases.  The only exception when using multiple threads is \texttt{dragon}.  The version in \cite{morozov2020lock} performs better and better on \texttt{dragon} when increasing the number of parallel resources, while ours (see Figure \ref{fig:scaling}) does not.  The main culprit is that, while in \cite{morozov2020lock} parallel resources are allocated only when needed in the computation, our thread pool (see Section \ref{subsec:Cpp_backend}) will allocate all the parallel resources indicated by the user ahead of time.  Our approach is most beneficial when the allocated resources can be reused during the computation, and this is true e.g.\ when computing homology dimensions in degree $2$ and above.  However, when computing only up to dimension $1$, it is only necessary to allocate the parallel resources once, and an on-the-fly approach such as the one in \cite{morozov2020lock} can be faster.  Another logically independent reason for this observed performance loss has to do with apparent pairs: we remind the reader that the implementation in \cite{morozov2020lock} is based upon \textit{Ripser v1.1} which, unlike \textit{Ripser v1.2} considered here, did not make use of the apparent pairs optimization.  While the search for apparent pairs and subsequent column assembly step is performed in parallel in homology dimension $1$ or higher, it is only done serially in dimension $0$.


\begin{figure}[H]
    \begin{minipage}[t]{.49\linewidth}
        \centering
        \includegraphics[width=\columnwidth]{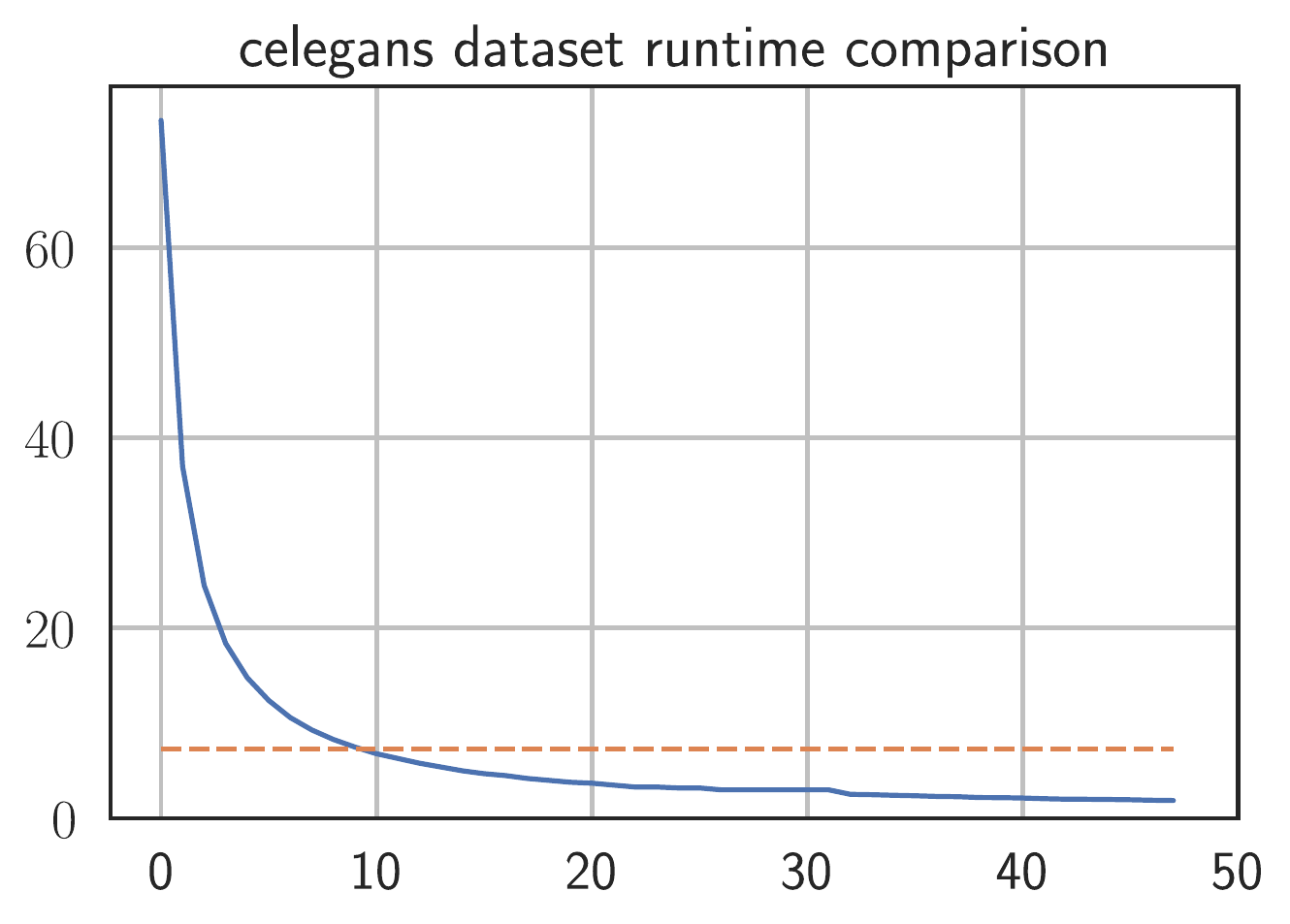}
        \label{fig:++_celegans}
    \end{minipage}
    \hfill
    \begin{minipage}[t]{.49\linewidth}
        \centering
        \includegraphics[width=\columnwidth]{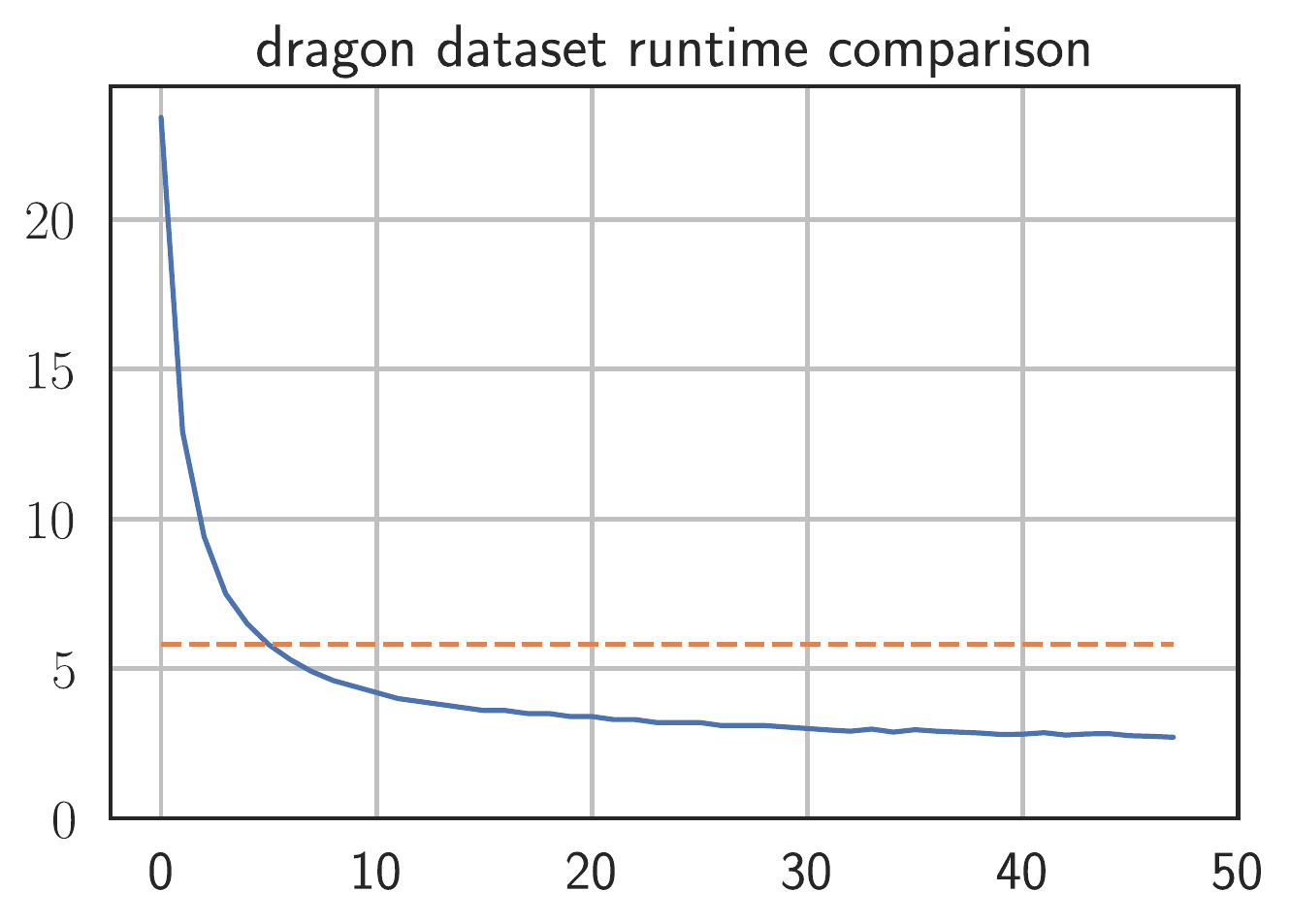}
        \label{fig:++_dragon}
    \end{minipage}
        \begin{minipage}[t]{.49\linewidth}
        \centering
        \includegraphics[width=\columnwidth]{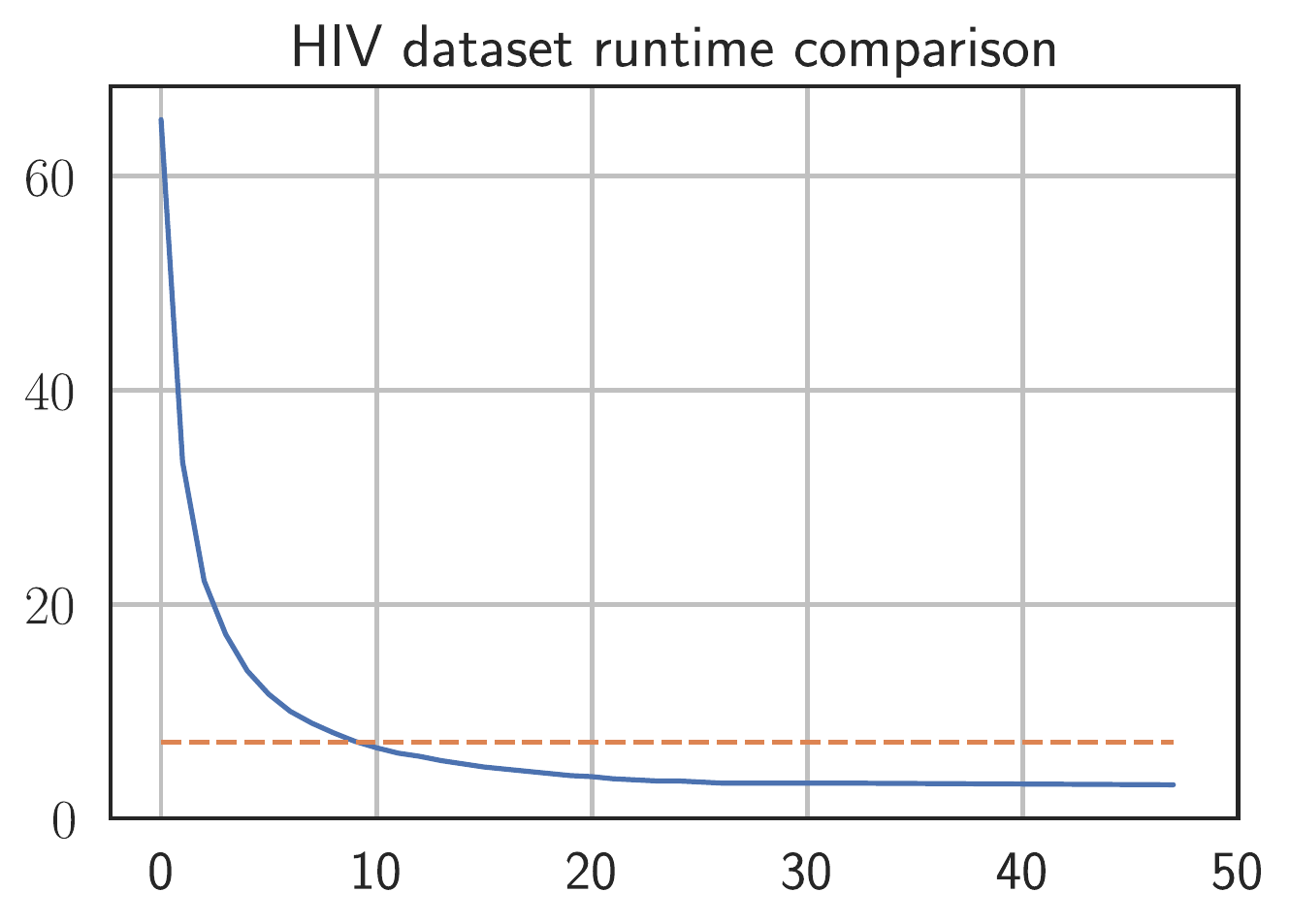}
        \label{fig:++_hiv}
    \end{minipage}
    \hfill
    \begin{minipage}[t]{.49\linewidth}
        \centering
        \includegraphics[width=\columnwidth]{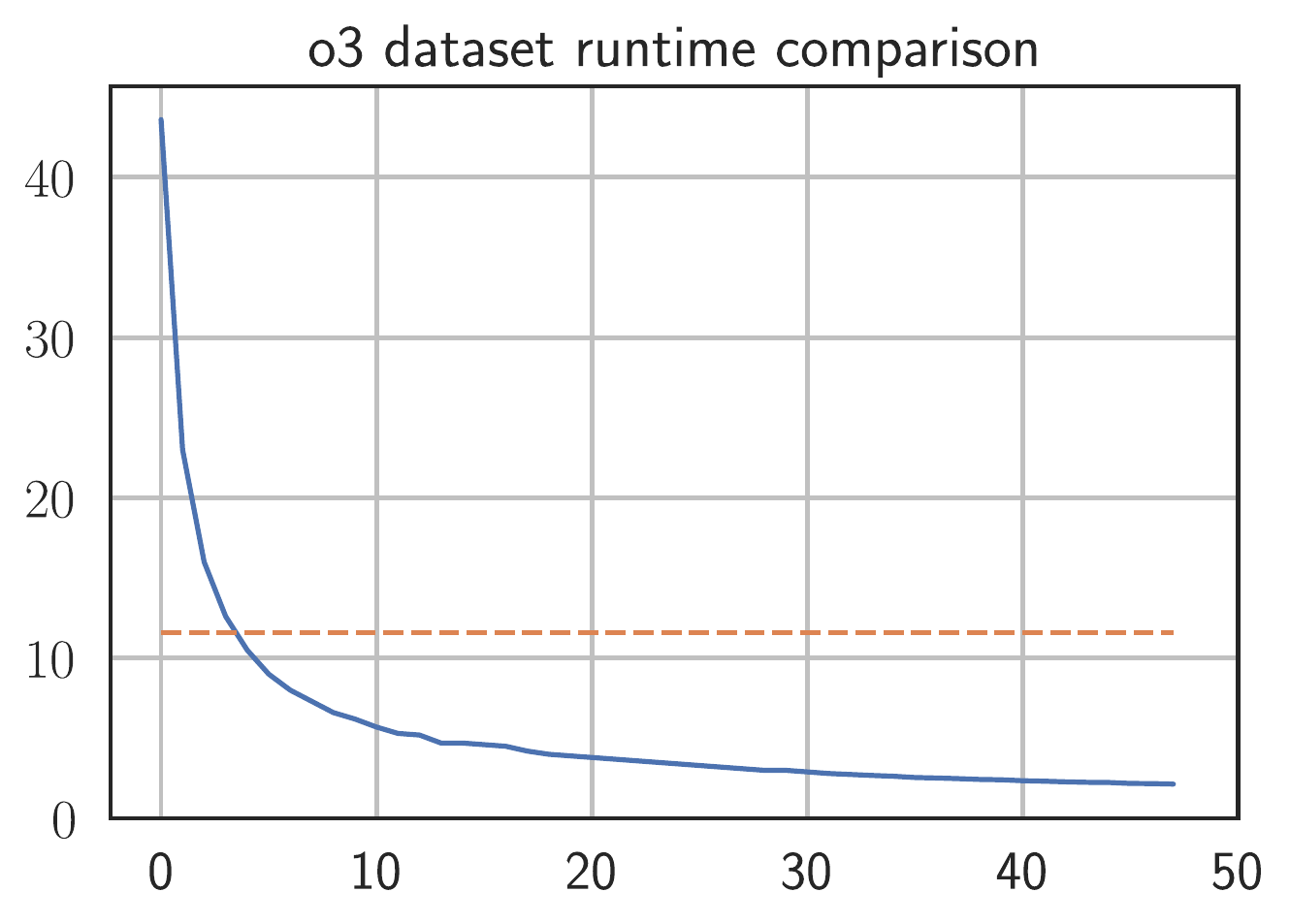}
        \label{fig:++_o3}
    \end{minipage}
        \begin{minipage}[t]{.49\linewidth}
        \centering
        \includegraphics[width=\columnwidth]{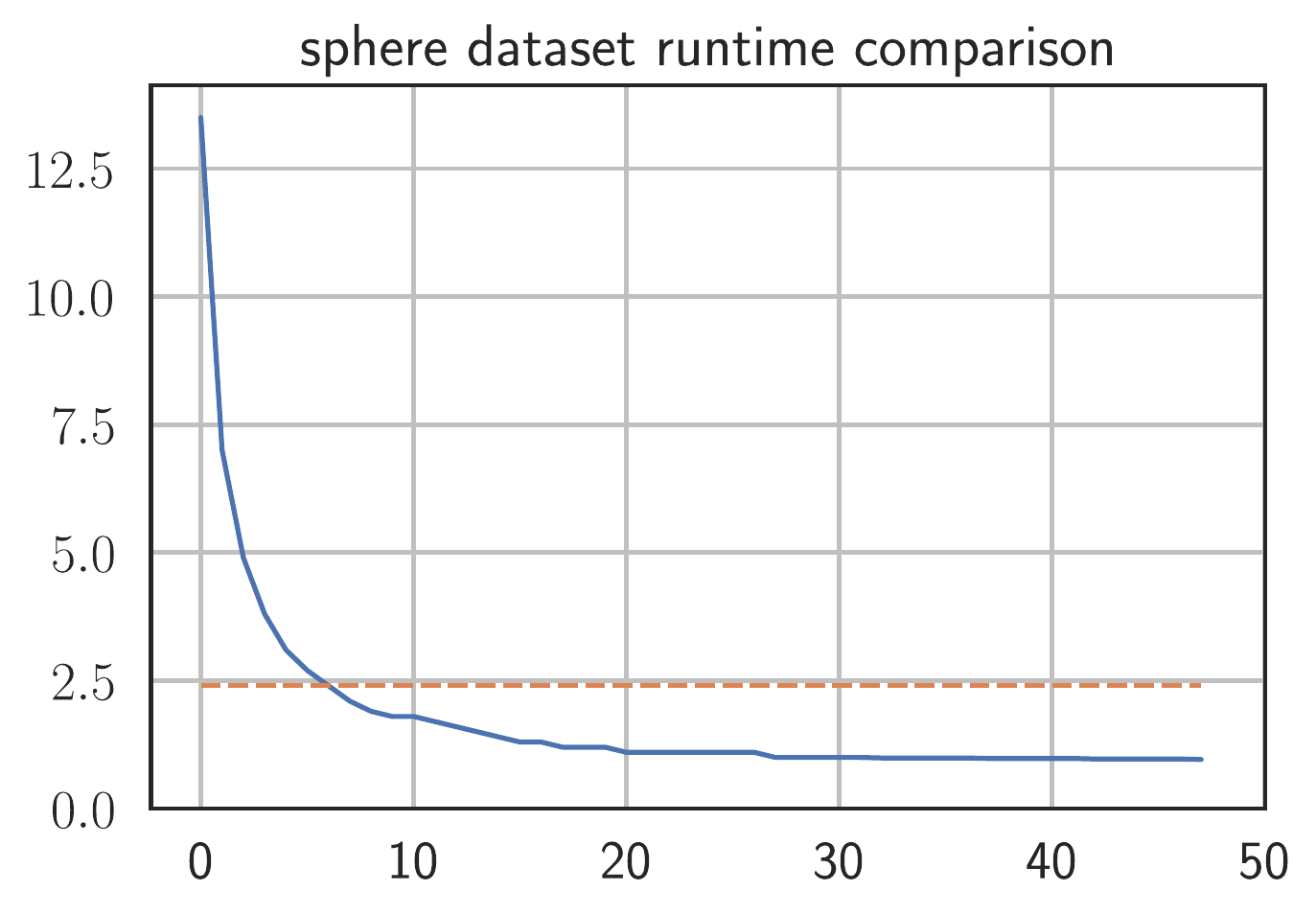}
        \label{fig:++_sphere}
    \end{minipage}
    \hfill
    \begin{minipage}[t]{.49\linewidth}
        \centering
        \includegraphics[width=\columnwidth]{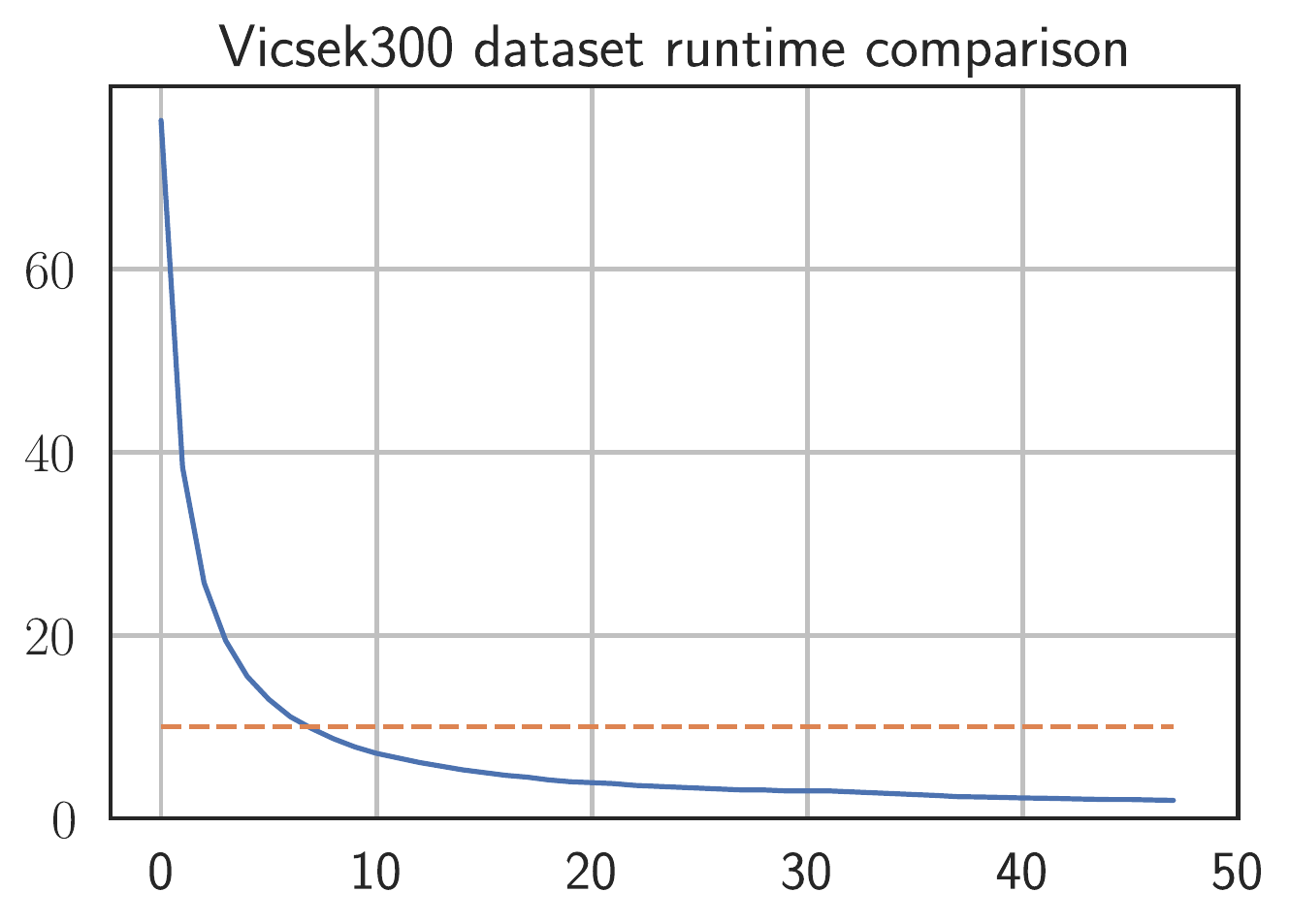}
        \label{fig:++_vicsek}
    \end{minipage}
    \caption{Run times comparison of \textit{giotto-ph} (full blue line) and \textit{Ripser++} (dashed orange line) using datasets from \cite{zhang2020gpuaccelerated}. The $x$-axis represent the number of threads used and the $y$-axis  the time (in seconds) to complete the PH computation.}
    \label{fig:comparison_gph_rpp}
\end{figure}

Considering the good performance obtained, we decided to compare our implementation with the state-of-the-art parallel code running on GPU: \textit{Ripser++} \cite{zhang2020gpuaccelerated}. For this test, we ran our code on the same datasets used in \cite{zhang2020gpuaccelerated} (for full details check Table 2 on page 23 of \cite{zhang2020gpuaccelerated}) and compared our run-times with the reported figures.
Figure \ref{fig:comparison_gph_rpp} shows that on our test machine, we achieve better performance when using only 4 to 10 threads, depending on the dataset, confirming that a relatively new CPU with at least 8 cores should be able to beat a high end GPU on this computation. 

There are multiple limitations in \textit{Ripser++} that were addressed in \textit{giotto-ph}. First, \textit{Ripser++} does not perform the matrix reduction in parallel. Second, apparent pairs are stored in a sorted array in order to provide apparent pair lookups in $\mathcal{O}(\log{}n)$ time using binary searches. Since it is possible to carry out the matrix reduction without recording and/or sorting apparent pairs, \textit{giotto-ph} results in a competitive solution, even if running on less high-performance hardware.



\begin{figure}[h!]
    \begin{center}
        \includegraphics[width=0.8\linewidth]{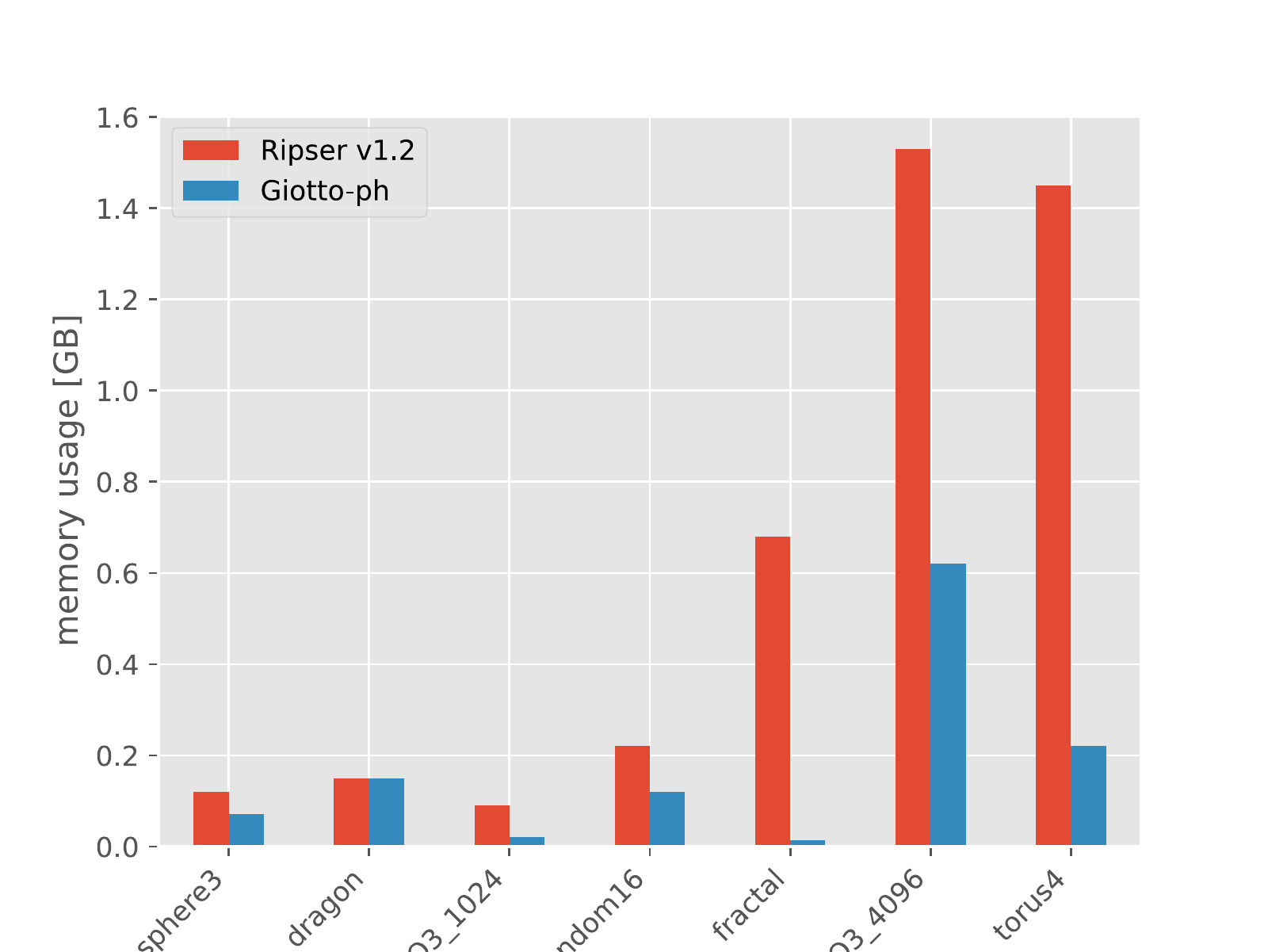}
    \end{center}
    \caption{Memory consumption of \textit{giotto-ph} and \textit{Ripser v1.2}. For the \texttt{dragon} dataset, the memory consumption is exactly the same because  the memory optimization only kicks in from dimension $2$.}
    \label{fig:mem}
\end{figure}

Figure \ref{fig:mem} compares the memory consumption of \textit{giotto-ph} and \textit{Ripser v1.2}.  Since these numbers are quite remarkable, we investigated the source of these differences and discovered that \cite{bauer2021ripser} missed an optimization when the required dimension is greater than $2$. We contributed with a pull request\footnote{\url{https://github.com/Ripser/ripser/pull/37}} to \cite{bauer2021ripser} that, once accepted, will make the memory benefit of our solution vanish.


\subsection{Higher homology dimensions}
\label{subsec:dim}
Table \ref{tab:higher} compares \textit{Ripser v1.2} and \textit{giotto-ph} when increasing the homology dimension parameter.  We included the measurements using EC to show the potential benefits. It is important to note that timings reported using EC do not include EC processing time; the interested reader can find them in Table \ref{tab:collapser}. The first dimension reported in Table \ref{tab:higher} is the one in the Table \ref{tab:datasets} setup.

\texttt{sphere3} and \texttt{random16} are the only datasets where Maximal Index (\textbf{MI})\footnote{\label{foot:MI}\textbf{MI} is the maximum number of retrievable entries in a data structure.} is not attained. \texttt{sphere3} is a highly regular dataset and computing higher homology dimensions will not yield interesting results.  \texttt{random16} produces no barcodes at dimension $20$. We arbitrarily decided to stop at dimension $10$ and report the data.

Table \ref{tab:datasets} shows that, in general, pre-processing with EC leads to a reduction in later run-times.  The only exception is the \texttt{sphere3} dataset, where EC is slightly detrimental.  The reason for this is implementational in nature as we now explain.  The EC step takes as input the dataset's distance matrix in dense format, and outputs a sparse matrix.  In the case of \texttt{sphere3}, EC removes very few edges, producing a highly filled sparse matrix.  Dense representations have better cache behaviour than sparse ones, and thus can lead to faster computations than highly filled sparse ones.  We are working on an heuristic to automatically select the best data format.


\begin{table*}
\footnotesize
    \begin{subtable}[t]{0.51\linewidth}
        \centering
        \begin{tabularx}{\linewidth}{|X|r|r|c|c|}
        \hline
        \multicolumn{5}{|c|}{\texttt{sphere3}} \\ \hline
        & \multicolumn{1}{c|}{\textbf{2}} & \multicolumn{1}{c|}{\textbf{3}} & \textbf{4} & \textbf{5} \\ \hline
        \textit{Ripser}                & 0.7 & 11.6 & 392 & \textbf{OOM}              \\ \hline
        \textit{Ripser} after EC    & 0.8 & 18.6 & 658 & \textbf{OOM}              \\ \hline
        \textit{giotto-ph}             & 0.4 & 2.2 & \multicolumn{1}{r|}{63} & \multicolumn{1}{r|}{1826} \\ \hline
        \textit{giotto-ph} after EC & 0.4 & 3.3 & \multicolumn{1}{r|}{97} & \multicolumn{1}{r|}{2900} \\ \hline
        \end{tabularx}
        \label{fig:hig_sphere}
    \end{subtable}
    \hfill
    \begin{subtable}[t]{0.48\linewidth}
        \centering
        \begin{tabularx}{\linewidth}{|X|r|r|r|c|}
        \hline
        \multicolumn{5}{|c|}{\texttt{o3\_1024}}    \\ \hline
        & \multicolumn{1}{c|}{\textbf{3}} & \multicolumn{1}{c|}{\textbf{4}} & \multicolumn{1}{c|}{\textbf{5}} & \textbf{6}                                            \\ \hline
        \textit{Ripser}                & 2.3 & 9.9 & 34.2 &   \\ \cline{1-4}
        \textit{Ripser} after EC    & 0.2 & 0.3 & 0.4  &   \\ \cline{1-4}
        \textit{giotto-ph}      & 0.5  & 2.2 & 7.9  &   \\ \cline{1-4}
        \textit{giotto-ph} after EC & 0.1 & 0.1 & 0.1  &   \multirow{-4}{*}{\textbf{MI}} \\ \hline
        \end{tabularx}
        \label{fig:hig_o31024}
    \end{subtable}
    \begin{subtable}[t]{0.51\linewidth}
        \centering
        \begin{tabularx}{\linewidth}{|X|r|r|c|}
        \hline
        \multicolumn{4}{|c|}{\texttt{o3\_4096}} \\ \hline
        & \multicolumn{1}{c|}{\textbf{3}} & \multicolumn{1}{c|}{\textbf{4}} & \textbf{5}                                            \\ \hline
        \textit{Ripser}                 & 45.6 & 334 &  \\ \cline{1-3}
        \textit{Ripser} after EC     & 1.9 & 3.6  &  \\ \cline{1-3}
        \textit{giotto-ph}              & 9.3 & 69   &  \\ \cline{1-3}
        \textit{giotto-ph} after EC  & 0.5 & 0.9  & \multirow{-4}{*}{\textbf{MI}}   \\ \hline
        \end{tabularx}
        \label{fig:hig_o34096}
    \end{subtable}
    \hfill
    \begin{subtable}[t]{0.48\linewidth}
        \centering
        \begin{tabularx}{\linewidth}{|X|r|r|r|r|}
        \hline
        \multicolumn{5}{|c|}{\texttt{random16}} \\ \hline
        & \multicolumn{1}{c|}{\textbf{7}} & \multicolumn{1}{c|}{\textbf{8}} & \multicolumn{1}{c|}{\textbf{9}} & \multicolumn{1}{c|}{\textbf{10}}                        \\ \hline
        \textit{Ripser}                & 3.9 & 7.9 & 13.6 & 20.3 \\ \hline
        \textit{Ripser} after EC   & 0.1 & 0.1 & 0.1 & 0.1 \\ \hline
        \textit{giotto-ph}             & 1 & 2.2 & 3.9 & 5.9 \\ \hline
        \textit{giotto-ph} after EC & 0.1 & 0.1 & 0.1 & 0.1 \\ \hline
        \end{tabularx}
        \label{fig:hig_random}
    \end{subtable}
    \begin{subtable}[t]{\linewidth}
        \centering
        \begin{tabularx}{\linewidth}{|X|r|r|c|c|r|r|c|}
        \hline
        \multicolumn{8}{|c|}{\texttt{fractal}} \\ \hline
        & \multicolumn{1}{c|}{\textbf{2}} & \multicolumn{1}{c|}{\textbf{3}} & \textbf{4}  & \textbf{5} & \multicolumn{1}{c|}{\textbf{6}} & \multicolumn{1}{c|}{\textbf{7}} & \textbf{8}                                            \\ \hline
        \textit{Ripser}                & 5.4 & \multicolumn{5}{c|}{\textbf{OOM}} &    \\ \cline{1-7}
        \textit{Ripser} after EC    & 0.1 & 0.2 & \multicolumn{1}{r|}{1} & \multicolumn{1}{r|}{3.4} & 10 & 23.5 &   \\ \cline{1-7}
        \textit{giotto-ph}             & 0.9 & 90 & 9100 & \multicolumn{3}{c|}{\textbf{OOM}} &   \\ \cline{1-7}
        \textit{giotto-ph} after EC & 0.1 & 0.1 & \multicolumn{1}{r|}{0.2} & \multicolumn{1}{r|}{0.9} & 2.5 & 6 &   \multirow{-4}{*}{\textbf{MI}} \\ \hline
        \end{tabularx}
        \label{fig:hig_fractal}
    \end{subtable}
    \begin{subtable}[t]{\linewidth}
        \centering
        \begin{tabularx}{\linewidth}{|X|r|r|c|c|c|}
        \hline
        \multicolumn{6}{|c|}{\texttt{dragon}}  \\ \hline
        & \multicolumn{1}{c|}{\textbf{1}} & \multicolumn{1}{c|}{\textbf{2}} & \textbf{3} & \textbf{4} & \textbf{5} \\ \hline
        \textit{Ripser}                & 2.2 & \multicolumn{1}{c|}{168} & \multicolumn{2}{c|}{\textbf{OOM}} &  \\ \cline{1-5}
        \textit{Ripser} after EC    & 0.1 & 0.9 & \multicolumn{1}{r|}{8.6} & \multicolumn{1}{r|}{85} &   \\ \cline{1-5}
        \textit{giotto-ph}             & 1.4 & 40 & 6508  & \textbf{OOM} &   \\ \cline{1-5}
        \textit{giotto-ph} after EC & 0.1 & 0.3 & \multicolumn{1}{r|}{1.8} & \multicolumn{1}{r|}{18} &   \multirow{-4}{*}{\textbf{MI}} \\ \hline
        \end{tabularx}
        \label{fig:hig_dragon}
    \end{subtable}
    \caption{Timings in seconds of \textit{Ripser v1.2} and \textit{giotto-ph} when increasing the maximal homology dimension to compute. \textit{giotto-ph} was run using $8$ threads. For both pieces of software, we included the run-times when using the output of EC as input to the respective persistent homology backends, but the timings do not include EC computation time.  \textbf{OOM} stands for ``out of memory'' while \textbf{MI} is the Maximal Index (see Footnote \ref{foot:MI}). The \texttt{torus4} dataset is not present here because MI is reached when the maximum homology dimension is set to $3$.} 

    \label{tab:higher}
\end{table*}


\subsection{Edge Collapser}
\label{subsec:Collapser}

We now report experimental findings concerning our EC implementation.  These are summarized in Table \ref{tab:collapser}, where the third column demonstrates that our solution is always faster than \textit{GUDHI}'s original one on the datasets considered.

We remind the reader that, as explained in Section \ref{sec:python}, a novelty of our implementation is the use of the enclosing radius computation to shorten the run-time of the EC step even beyond what is already made possible by our use of faster routines and data structures.  The experimental impact of this enhancement is shown in the last column of Table \ref{tab:collapser}.  One would expect that the more ``random'' datasets, where ``central points'' are likely to be present, will benefit the most from thresholding by the enclosing radius.  Among our standard datasets from Table \ref{tab:datasets}, \texttt{random16}, \texttt{o3\_1024} and \texttt{o3\_4096} are random datasets, but we do not witness such an impact.  While, in the case of \texttt{random16}, the reason is likely that the dataset it too small ($50$ points),  in the case of the \texttt{o3} datasets the reason is that a threshold lower than the enclosing radius is provided, meaning that the enclosing radius optimization is not used at all there.  To demonstrate that our expectation is valid despite the limitations caused by our choice of datasets and configurations, we have added an entry to Table \ref{tab:datasets}, representing a dataset of $3000$ points sampled from the uniform distribution on the unit cube in $\mathbb{R}^3$.  Together with \texttt{dragon}, this example shows that large gains can be made by using the enclosing radius on certain datasets.

\begin{table}[h!]
\centering
\begin{tabular}{|l|r|r|r|r|}
\hline
\multicolumn{1}{|p{2cm}|}{\textbf{dataset}} & {\textbf{\textit{GUDHI} EC}} & \multicolumn{1}{p{2.4cm}|}{\textbf{\textit{giotto-ph} EC} (speedup)} & \multicolumn{1}{p{2.5cm}|}{\textbf{\textit{giotto-ph} EC with encl.\ rad.} (speedup)} \\ \hline
\texttt{sphere3}   & 1.6    & 0.9 (1.78)   & 0.9 (1.78)   \\ \hline
\texttt{dragon}    & 63     & 36 (1.75)    & 28 (2.25)   \\ \hline
\texttt{o3\_1024}  & 0.2    & 0.13 (1.53)  & 0.13* (1.53) \\ \hline
\texttt{random16}  & 0.004  & 0.001 (4.00)    & 0.001 (4.00) \\ \hline
\texttt{fractal}   & 1.32   & 0.8 (1.65)   & 0.8 (1.65)   \\ \hline
\texttt{o3\_4096}  & 2.1    & 1.2 (1.75)   & 1.2* (1.75)  \\ \hline
\texttt{torus4}    & 10     & 6.7 (1.49)    & 6.7* (1.49)   \\ \hhline{|=|=|=|=|}
\multicolumn{1}{|p{1.7cm}|}{3000 pts in unit cube}    & 180    & 125 (1.44)    & 78 (2.31) \\ \hline
\end{tabular}
\caption{Run-time comparison between \textit{GUDHI}'s implementation \cite{gudhi:Collapse} of the EC algorithm of Boissonnat and Pritam \cite{boissonnat2020edge} and \textit{giotto-ph}'s implementation.  The last column reports run-times when sparsifying according to the enclosing radius before calling \textit{giotto-ph}'s EC, which is the default behaviour when no threshold is provided by the user.  All execution times are in seconds, while speedups are ratios.  Cells marked with an asterisk mean that a threshold is provided and therefore the enclosing radius is not computed by default. The last entry is unique to this table and better demonstrates the impact of the enclosing radius optimization on favourable datasets and configurations.}
\label{tab:collapser}
\end{table}

\subsection{Low-end CPU}

\begin{figure}[H]
    \begin{center}
        \includegraphics[width=0.8\linewidth]{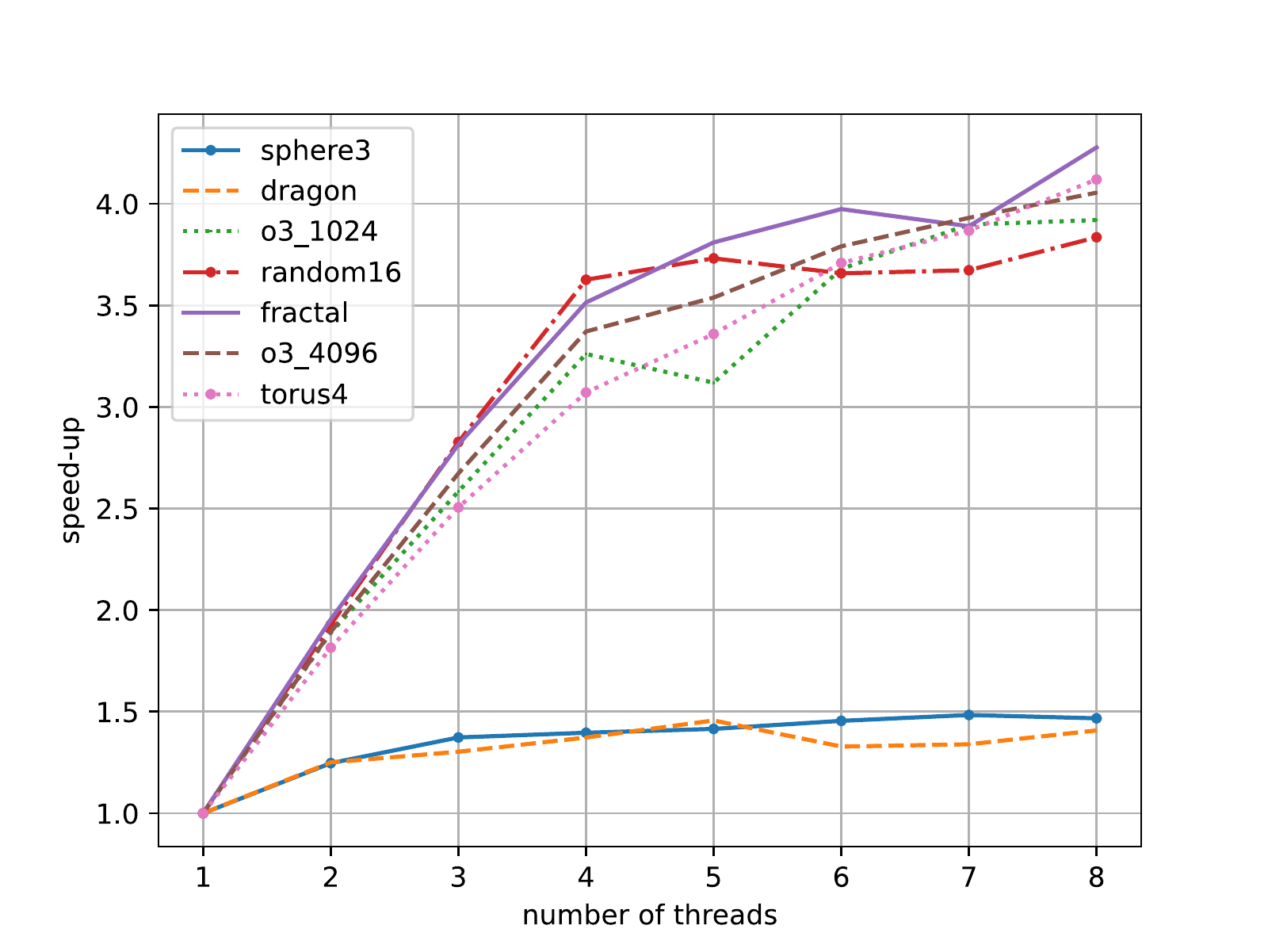}
    \end{center}
    \caption{Scaling of \textit{giotto-ph}'s PH backend on a low-end CPU. \textit{giotto-ph} uses $8$ threads fully exploiting the hardware resources: $4$ physical cores and $8$ logical threads (``hyper-threading'').}
    \label{fig:cpu}
\end{figure}

Not all researchers have access to high-end machines. In this section we report performance figures when using a low-mid end CPU, showing that similar results can be achieved. The CPU used in this test is an Intel\textsuperscript{\textregistered} Core\textsuperscript{\texttrademark} i7-7700 CPU with 4 physical cores. Figure \ref{fig:cpu} shows how the scaling of \textit{giotto-ph}'s PH backend with the number of threads, when using this CPU.  Comparing this plot to Figure \ref{fig:1.2}, one sees that the performance figures are very similar despite the use of a less high-performing hardware platform.

\section{Conclusion and future work}
\label{sec:conclusion}

We integrated multiple, existing and novel, algorithmic ideas to obtain a state-of-the-art implementation of the computation of persistent homology for Vietoris--Rips filtrations.  This implementation enables the use of parallel CPU resources to speed up the computation and outperforms even state-of-the-art GPU implementations.

We plan to extend \textit{giotto-ph} by supporting a wider range of filtrations in a modular way.  We also plan to add features (e.g.,\ simplex pairs and essential simplices) needed for back-propagation in a deep learning context, and seamless integration with frameworks such as \textit{PyTorch}.\footnote{\url{https://pytorch.org/}}


\section*{Acknowledgements}
The authors would like to thank Kathryn Hess Bellwald for numerous fruitful discussions as well as Ulrich Bauer for very helpful conversations about \textit{Ripser}. This work was supported by the Swiss Innovation Agency (Innosuisse project 41665.1 IP-ICT).


\bibliography{bibliography}
\end{document}